# Local Alloy Order in a $Ge_{1-x}Sn_x$/Ge Epitaxial Layer

Agnieszka Anna Corley-Wiciak[1,2], Shunda Chen[3*], Omar Concepción[4], Marvin Hartwig Zoellner[1], Detlev Grützmacher[4], Dan Buca[4], Tianshu Li[3], Giovanni Capellini[1,5], Davide Spirito[1*]

[1] IHP – Leibniz-Institut für innovative Mikroelektronik, 15236 Frankfurt (Oder), Germany

[2] RWTH Aachen University, 52056 Aachen, Germany

[3] Department of Civil and Environmental Engineering, George Washington University, Washington, DC 20052, USA

[4] Peter Grünberg Institute 9 (PGI-9) and JARA-Fundamentals of Future Information Technologies, 52428 Jülich, Germany

[5] Dipartimento di Scienze, Università Roma Tre, V.le G. Marconi 446, 00146 Roma, Italy

* shdchen@email.gwu.edu, spirito@ihp-microelectronics.com

**Abstract**

The local ordering of atoms in alloys directly has a strong impact on their electronic and optical properties. This is particularly relevant in nonrandom alloys, especially if they are deposited using far from the equilibrium processes, as is the case of epitaxial $Ge_{1-x}Sn_x$ layers. In this work, we investigate the arrangement of Ge and Sn atoms in optoelectronic grade $Ge_{1-x}Sn_x$ epitaxial layers featuring a Sn content in the 5-14% range by using polarization-dependent Raman spectroscopy and density-functional-theory calculations. The thorough analysis of the polarization-dependent spectra in parallel and perpendicular configuration allowed us to properly tag all the observed vibrational modes, and to shed light on that associated to disorder-assisted Raman transitions. Indeed, with the help of large-scale atomistic simulations, we were able to highlight how the presence of Sn atoms, that modify the local environments of Ge atoms, gives rise to two spectral features at different Raman shifts, corresponding to distortions of the atomic bonds. This analysis provides a valuable framework for advancing the understanding of the vibrational properties in $Ge_{1-x}Sn_x$ alloys, particularly with regard to the impact of local ordering of the different atomic species.

## I. Introduction

Germanium-tin (Ge$_{1-x}$Sn$_x$) semiconductor alloys have been successfully used in the fabrication of photonic and nanoelectronic devices and are deemed promising for other applications such as energy harvesting [1–8]. Their electronic conduction and valence bands can be engineered to offer an indirect or fundamental direct bandgap [9–11]: a direct band gap is expected for a fully relaxed layer having a Sn content of $x \gtrsim 8\%$ [12–18], but tensile and compressive biaxial strain can decrease or increase, respectively, this value. However, the technological development of these CMOS-compatible alloys faces several obstacles related to the very low solid solubility and the larger atomic mass of Sn [19]. Consequently, growth conditions far away from thermodynamic equilibrium are needed to overcome the solubility limit. Indeed, Sn atoms tend to aggregate, negatively impacting optical and electrical properties. Even when large Sn clusters are not present, alloy configuration different from the classical random alloy assumption (i.e., if each neighbor site is not randomly occupied according to the composition) will significantly affect the materials properties, as the overlap of the atomic orbitals controls the resulting band structure [20,21]. In this respect, Ge$_{1-x}$Sn$_x$ alloys differ from silicon-germanium, where random alloy is a valid approximation [20,22]. Numerical and experimental results indicated the presence of a short-range order (SRO) in Ge$_{1-x}$Sn$_x$, with a reduction of the coordination number of the Sn (i.e., a repulsion of among Sn pairs) [20,23–25], which should be observable above a Sn content of 10%, a composition just above that required for a direct bandgap alloy. In general, the local environment of Ge atoms in Ge$_{1-x}$Sn$_x$ alloy, even at low Sn content, is distorted with respect to bulk Ge; additionally, it is possible that the thermodynamically stable configuration contains multiple environments with different levels of distortion, which would affect the vibrational, optical and electrical properties of the material [20,21].

Experimentally, methods sensitive to atomic scale, such as Atomic Probe Tomography (APT) [26] or Extended X-ray absorption fine structure (EXAFS) [23–25] are often used. Unfortunately, these methods require large facilities and/or complex and destructive sample preparation. As such, an easy-to-access, high-throughput, non-destructive experimental method enabling the investigation of atomic ordering is highly welcome. Raman spectroscopy is a suitable choice, as it provides detailed information on local vibrational modes. In cubic group-IV elemental semiconductors with diamond structure, such as Si and Ge, the Raman spectrum consists of a single, well-defined peak associated with first-order scattering, and weaker, convoluted second-order modes. In Ge$_{1-x}$Sn$_x$ alloys multiple first-order peaks are expected, associated with vibrations of Ge-Ge, Ge-Sn and Sn-Sn pairs, as well as spectral features due to the relaxation of selection rules in a disordered alloy. The analysis of the spectra can be further helped by polarization resolution, which allows multiple modes to be deconvolved according to their symmetry. This method is very efficiently complemented by first-principles density-functional-theory (DFT) calculations of the vibrational modes under different disorder configurations.

In this work, a large set of epitaxial $Ge_{1-x}Sn_x$ layers are investigated. The layers are grown in an industry-grade chemical vapor deposition (CVD) reactor on a Ge/Si virtual substrate [6]. Layer thicknesses and build-in biaxial strains are reported in Appendix A. A detailed Raman-polarized spectroscopy study is presented for two $Ge_{1-x}Sn_x$ layers containing 6 and 14 at. % Sn. DFT simulations are used to assign the spectral features to specific vibration in the alloy. Next, the effect of composition is addressed.

## II. Polarization dependence and local alloy ordering

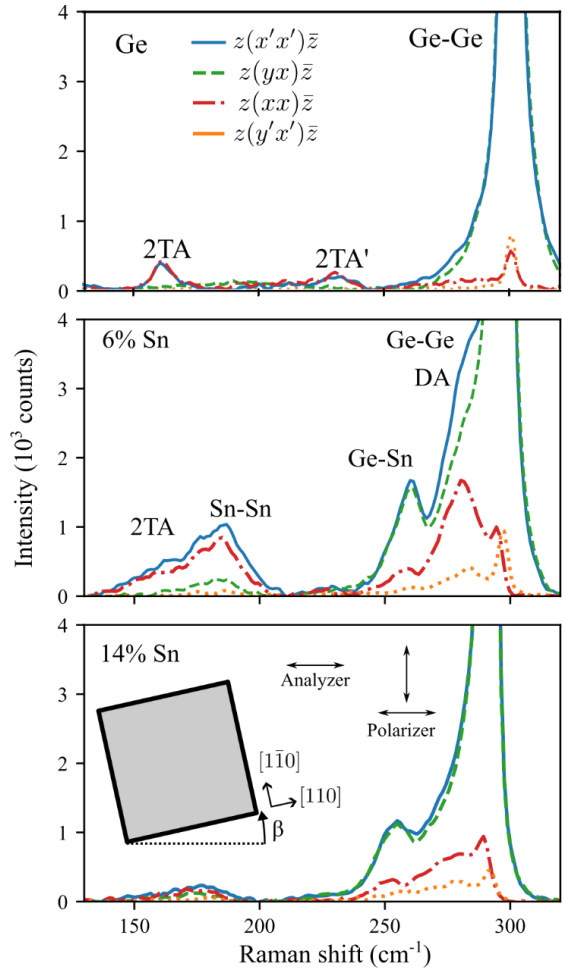

FIG. 1. Raman spectra for Ge (upper panel) and samples with 6at.% of Sn and strain of -0.32% (middle panel) and with 14at.% of Sn and strain of -0.51% (lower panel). The inset shows the measurement setup. The polarization configuration is indicated using Porto notation, with x=[100], y=[010], x'=[110], y'=[1$\bar{1}$0].

Raman spectroscopy was performed using an excitation laser at a wavelength of 633 nm in backscattering geometry. The setup has a fixed polarization analyzer at the entrance of the spectrometer, and a polarizer for the excitation laser that can be oriented parallel or perpendicular to the analyzer. The sample can be rotated with respect to these directions by an angle $\beta$, where $\beta = 0$ is the direction of the analyzer (see inset in FIG. 1). The laser is incident along the growth direction [001] ($z$).

The spectra acquired for specific polarization and alignment of the sample, according to the Porto notation with the definitions referred to crystal directions $x=[100], y=[010], x'=[110], y'=[1\bar{1}0]$ (see Table I) is shown in FIG. 1. The samples are mounted so that $x'$ is aligned to $\beta=0$.

For group-IV cubic semiconductors with diamond structure ($O_h$, $m\bar{3}m$ point group), the first-order Raman spectrum in backscattering geometry features a single peak, corresponding to longitudinal optical (LO) phonons belonging to $T_{2g}$ representation, and observable only in the polarization configurations $z(yx)\bar{z}$ and $z(x'x')\bar{z}$. In the case of alloys, as the symmetry is fulfilled only approximately, the selection rules for polarization can be less strict. To simplify the following discussion, we use the Raman-active representation of the $O_h$ point group as a guideline to study the polarization angle dependence of the modes in Ge$_{1-x}$Sn$_x$ and label Raman-active representations accordingly, i.e. $T_{2g}$, non-degenerate $A_{1g}$, and twofold degenerate $E_g$. The active representations in every configuration are reported in Table I.

*Table I Polarization configurations and active representations.* $e_i$ *is the polarization of the incident light,* $e_s$ *is the polarization at the analyzer.*

| Porto notation | Geometry | Active representations |
|---|---|---|
| $z(x'x')\bar{z}$ | $e_i \parallel \langle 110 \rangle$<br>$e_s \parallel \langle 110 \rangle$ | $A_{1g} + E_g + T_{2g}(LO)$ |
| $z(y'x')\bar{z}$ | $e_i \parallel \langle 1\bar{1}0 \rangle$<br>$e_s \parallel \langle 110 \rangle$ | $E_g$ |
| $z(xx)\bar{z}$ | $e_i \parallel \langle 100 \rangle$<br>$e_s \parallel \langle 100 \rangle$ | $A_{1g} + E_g$ |
| $z(yx)\bar{z}$ | $e_i \parallel \langle 010 \rangle$<br>$e_s \parallel \langle 100 \rangle$ | $T_{2g}(LO)$ |

The Raman spectra obtained for a Ge and Ge$_{1-x}$Sn$_x$ with 6 at.% Sn and a 14 at.% Sn is shown in Fig. 1. In case of pure Ge sample we can observe main LO mode at 300 cm$^{-1}$ and different 2$^{nd}$ order modes noted as 2TA (160cm$^{-1}$) and 2TA' (230cm$^{-1}$) [27]. For Ge$_{1-x}$Sn$_x$ the spectra acquired in LO-allowed configurations presents an intense peak at approximately 300 cm$^{-1}$, associated with Ge-Ge vibration mode. On the low energy side, the peak approximately 255 cm$^{-1}$ corresponds to the Ge-Sn vibration mode. In between these two modes, a spectral feature is present, labeled in literature as disorder-assisted (DA) [28–31]. It is usually argued that the DA originates from the relaxation of the momentum selection rule due to alloy disorder [32–38]. As expected, the Ge-Ge mode is strongly suppressed in the LO-forbidden configurations where the DA is relatively intense, indicating that it does not share the same symmetry. At lower energy of approximately 180 cm$^{-1}$, a spectral feature composed of multiple peaks without a clear polarization dependence appears. In literature, this is sometimes attributed to Sn-Sn pair vibration mode, although its anomalous polarization-behavior has already been noticed (it should follow the LO-allowed conditions), and is superimposed to the second-order contribution from two transverse acoustic phonons in Ge (2TA) [29,38].

It is noticed that all the peaks for the 14 at.%Sn layers are redshifted with respect to their counterparts in the 6 at.% Sn sample. Moreover, the spectra of the two samples differ in the relative intensity of the minor features which are stronger for 6 at.%Sn, such as the Sn-Sn group and the DA in *z(xx)z̄* configuration.

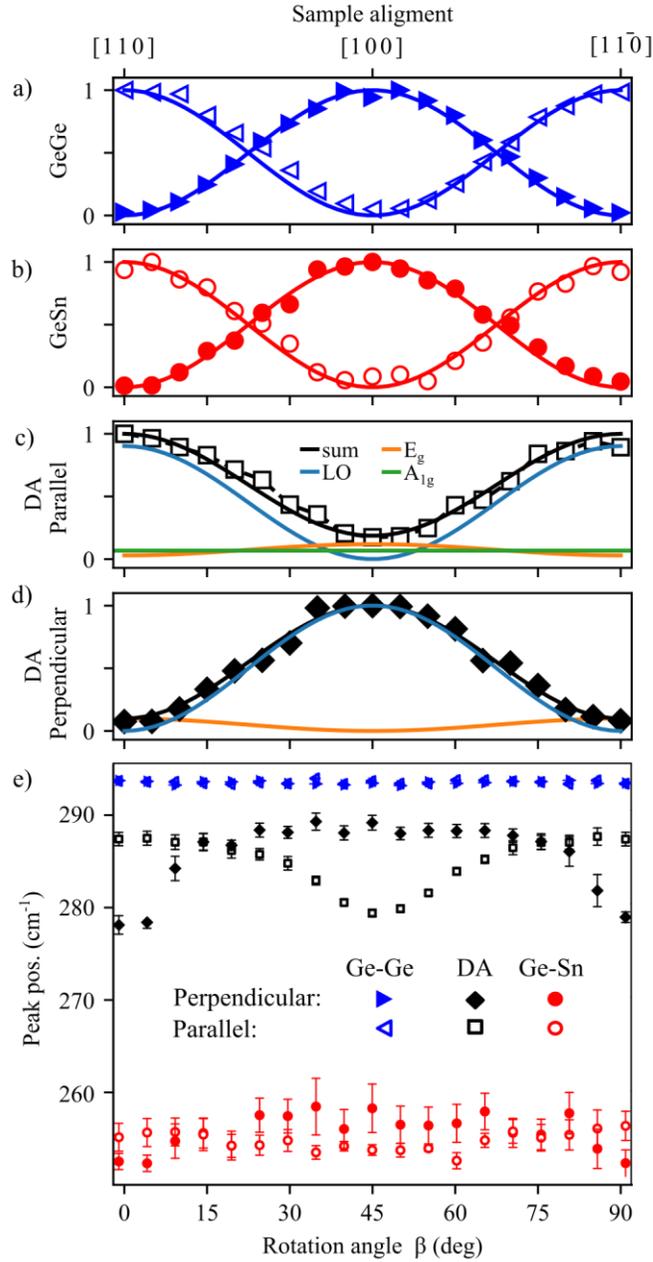

*FIG. 2. Angular dependence of the normalized intensity and position of the different peaks in the sample with 14 at.%, in the parallel (empty symbols) and perpendicular (full symbols) alignment of polarizer a),b) Intensity Ge-Ge and Ge-Sn vibration modes. The lines represent the fitting with an LO angular dependence. c),d) Fitting of the angular dependence of DA intensity with a combination of different representations. e) Angular dependence of the Raman position for Ge-Ge, DA, and Ge-Sn modes.*

To better illustrate this point, FIG. 2 presents the intensity and peak positions (obtained by the procedure described in Appendix C) vs the angle $\beta$ for the sample having the highest Sn content of 14 at.%. Data for both parallel and perpendicular alignment of the polarizer are reported here, with the intensity being normalized to its maximum value. The intensity shows an antiphase trend between parallel and perpendicular alignment, which is expected for a pure LO symmetry (see Appendix B). Regarding the peak positions, the following is noted: i) the Ge-Ge mode energy is independent of the angle and the polarization configuration; ii) the Ge-Sn mode energy has an angular dispersion of approximately 3 cm$^{-1}$ around the average value; iii) the DA peak is characterized by a significant angular dependency, correlated to the intensity variation. The former indicates that the DA peak is a convolution of multiple modes whose center of mass shifts with the angle because of their different symmetry.

The solid lines in Figs. 2(a) and 2(b) represent the expected angular dependence for a $T_{2g}$(LO) mode (formulae can be found in Appendix B) and fully fit the experimental data, for both the Ge-Ge and Ge-Sn modes. This is not the case for the DA peak (Figs. 2(c) and 2(d)). In particular, at $β=45°$, the mode intensity is significantly different from 0 in parallel configuration. The trend of the DA mode intensity can be explained by considering it as a sum of modes with different symmetry, comparing the angular dependence for the representations $E_g$ and $A_{1g}$, which have a maximum at $β=45°$. The DA mode has thus a component DA(LO) at approximately 289 cm$^{-1}$ and a component DA($A_{1g}$+$E_g$) at lower energy (approximately 280 cm$^{-1}$).

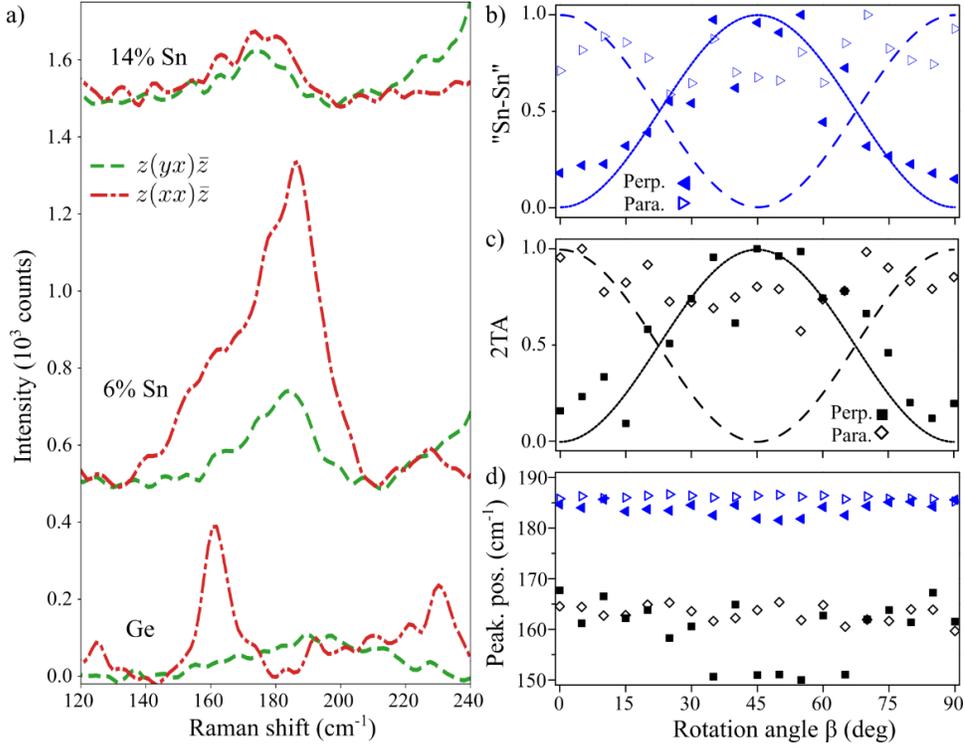

FIG. 3. a) Raman spectra for samples of bulk Ge, $Ge_{1-x}Sn_x$ with 6at.% of Sn and with 14at.% of Sn in the polarization configurations $z(xx)\bar{z}$ ($A_{1g} + E_g$) and $z(yx)\bar{z}$ (LO). Curves are offset for clarity. b,c) Angular dependence of peak intensity of "Sn-Sn" and "2TA" peaks, respectively, in the perpendicular (empty symbols) and parallel (full symbols) alignment for the sample with 6 at.% Sn, normalized to the maximum. Lines are the expected dependence for LO case for parallel (dashed) and perpendicular (solid) configurations. d) Angular dependence of the peak position for "Sn-Sn" and "2TA" in perpendicular and parallel polarization. Symbols are the same as in panels b,c.

A similar analysis was performed for the spectral feature observed at approximately 180 cm$^{-1}$ in the spectra acquired on the 6 at.% Sn sample, given its relatively higher intensity (see Fig. 3a). While bulk Ge, also shown in Fig. 3a, presents a single well-defined peak at 160 cm$^{-1}$, $Ge_{1-x}Sn_x$ samples have two peaks, one at approximately 165 cm$^{-1}$ and a stronger one at approximately 180 cm$^{-1}$, labeled "2TA" and "Sn-Sn" respectively (peak deconvolution is reported in Appendix C). The intensity of the two peaks does not follow, as a function of sample rotation angle β, the LO mode type behavior, as shown in Figs. 3(b) and 3(c). The 2TA mode of Ge should behave as an $A_{1g}$ mode, i.e. a constant amplitude in parallel and null amplitude in perpendicular. Instead, we observe an unclear, yet present, amplitude also in perpendicular alignment [39]. Consequently, these modes cannot be attributed to a simple Sn-Sn pair vibration, in agreement with its intensity independence of the Sn content, $x$. In fact, if it were related to the Sn-Sn vibration, an intensity proportional to $x^2$ would be expected (i.e., at 14 at.% Sn should be more than 5 times stronger than at 6 at.% Sn). As a combination of second-order modes it can be damped by the increased scattering with the disorder.

To better understand the origin of the spectral features, DFT calculations were performed to compute Raman spectrum of $Ge_{1-x}Sn_x$ alloys using the *Vienna Ab initio Simulation Package* (VASP) [40] based on the projector augmented wave method [40–42]. Local density approximation (LDA) [43] was employed for the exchange-correlation functional, which has been shown to yield the best agreement with experiments on elemental Ge and Sn for geometry optimization [20,44–46]. As discussed in Ref. [20], $Ge_{1-x}Sn_x$ alloys exhibit a non-random atomic distribution in the lattice through a short-range order (SRO) which can significantly affect electronic structures of the alloys for Sn content > 10 at.%. To understand the impact of atomic distribution on Raman modes, here we considered an SRO and a clustered-Sn structural model. A simulation cell containing 64 atoms, obtained by replicating a conventional diamond cubic cell containing eight atoms twice along each dimension, was chosen to ensure a sufficient sampling at the DFT level [20]. The SRO model was created through conducting the DFT-based Monte Carlo sampling at lattice temperature T=300 K [20]. A clustered-Sn model was created by replacing a cluster of 12 Ge atoms in the middle of a bulk Ge by Sn atoms, corresponding to a Sn content of 18.75%. Each structure undergoes full energy minimization where both the cell geometry and atomic positions are relaxed. The energy cutoff was chosen as 300 eV. To approach full optimization, the convergence for energy and force was set as $10^{-4}$ eV and $10^{-3}$ eV/Å, respectively. A 2 × 2 × 2 Monkhorst–Pack k-point grid was used to sample the first Brillouin zone [47]. The off-resonance Raman spectra are simulated using the formalism described by Porezag and Pederson [48] implemented in the VASP–Raman code and density functional perturbation theory (DFPT) [49,50]. An energy convergence threshold of $10^{-8}$ eV was used for DFPT calculations. The derivatives of the dielectric tensor were computed numerically by displacing the atoms 0.01 Å in both positive and negative directions. Note that since the simulations do not include the compressive lattice strain present in the experimental samples, and that LDA often results in over-binding, the calculated peak position can exhibit a small shift with respect to the experimental results. It is also noted that a higher content of Sn (18.75%) is selected to enable a focused study on the role of Sn on the DA peak.

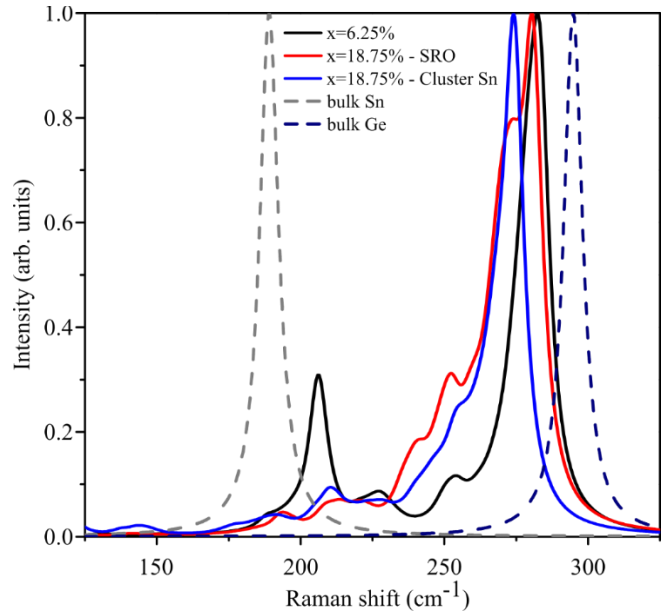

FIG. 4. DFT calculated Raman spectra for bulk Sn and Ge (dashed lines) and Ge$_{1-x}$Sn$_x$ with x~6% and x~19% in different ordering configurations (solid lines).

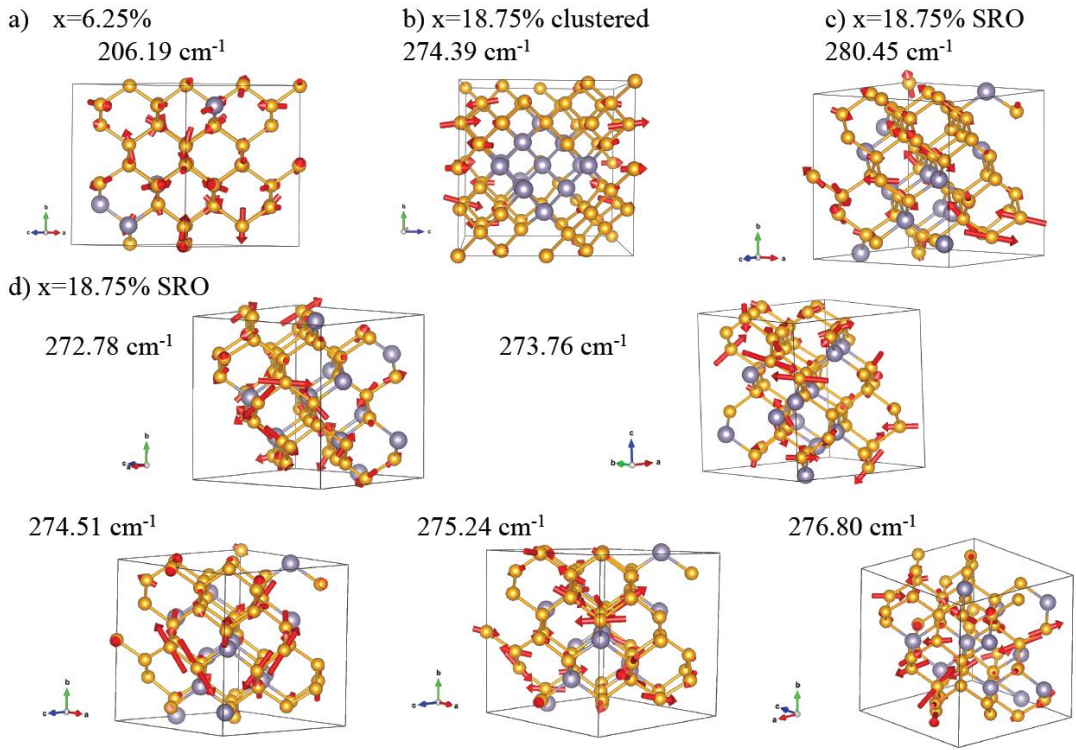

FIG. 5. a) Displacement vectors for the commonly assumed "Sn-Sn" peak for a low-Sn alloy model. b,c) Displacement vectors for the main peak for clustered and SRO Sn alloy models, respectively, d) Displacement vectors at different energy for the "DA shoulder" for a SRO alloy model with x~19%. Sn atoms are represented as gray balls, Ge atoms are orange.

As shown in Figs. 4 and 5, a shoulder is found to appear on the low energy side of the main Ge-Ge mode in the SRO configuration. In contrast, only a single main Ge-Ge mode (without shoulder) is present in the case of clustered Sn and it is shifted to a lower energy, matching the shoulder peak in the SRO case. The analysis of the atomic displacement at this main peak indicates that only Ge atoms contribute to the vibration (Fig. 5(b)), confirming the Ge-Ge nature of the main peak in clustered Sn, similar to the main Ge-Ge peak in the SRO case (Figs. 5(c) and 5(d)). The main Ge-Ge modes are further found to be attributed to the local Ge environment in $Ge_{1-x}Sn_x$ alloys resembling bulk Ge. Interestingly, the analysis indicates the shoulder mode appearing in the SRO case also only involves the vibration of Ge atoms, but for those within a local distorted environment induced by Sn-atoms. Thus the shoulder peak reflects the distribution of the Sn atoms to a certain extent. In this regard, the absence of the shoulder in the clustered Sn case is attributed to the fact that most Ge atoms share a very similar environment (the Sn being concentrated). Because the clustered Sn configuration is thermodynamically unfavored over SRO and has a greater lattice constant (longer Ge-Ge bond) [51], the Ge-Ge main peak of clustered Sn exhibits a redshift. Notably, because the shoulder peak observed in the SRO case matches the experimental DA peak qualitatively, we propose the DA peak to be assigned to Ge-Ge vibration affected by local lattice distortion caused by Sn atoms. Since lattice distortion within $Ge_{1-x}Sn_x$ is affected by how Sn atoms are distributed within the lattice, namely, being completely random or exhibiting an SRO, the DA($A_{1g}+E_g$) component at lower energy and with different symmetry can thus be interpreted as "genuine" disorder-activated scattering of phonons far from the zone center.

Shifting our attention to the lower energy "Sn-Sn" mode, we find all vibrations around its energy involve displacements of both Sn and Ge atoms. In fact, the "Sn-Sn" mode involves vibrations participated mainly from Ge atoms (as shown in the additional simulations shown in Appendix D). This observation is quantified by counting the contribution of each species to the mode. For the SRO configuration of approximately 19% Sn alloy, 52 atoms vibrate in correspondence with the mode at 193 $cm^{-1}$, of which 43 are identified to be Ge atoms. For a Sn content of 6%, 53 Ge atoms out of 56 atoms are found to contribute to this mode. The fact that this proportion follows the composition of Ge indicates that these low-energy modes cannot be associated with specific pairs. Considering their experimental polarization dependence that does not match the expected $T_{2g}$(LO) of the first-order nature, the modes are subsequently assigned to second-order scattering., analogous to that observed in the Ge sample. Different from bulk Ge, which has a single peak in this region (Fig. 3(a)), the two peaks observed in $Ge_{1-x}Sn_x$ alloys can be originated from a splitting in the modes at the zone edge, or from the difference of local environments in analogy to the Ge-Ge peak. Additionally, both in experiment and simulation, the modes are damped as the Sn content increases.

## III. Composition dependence

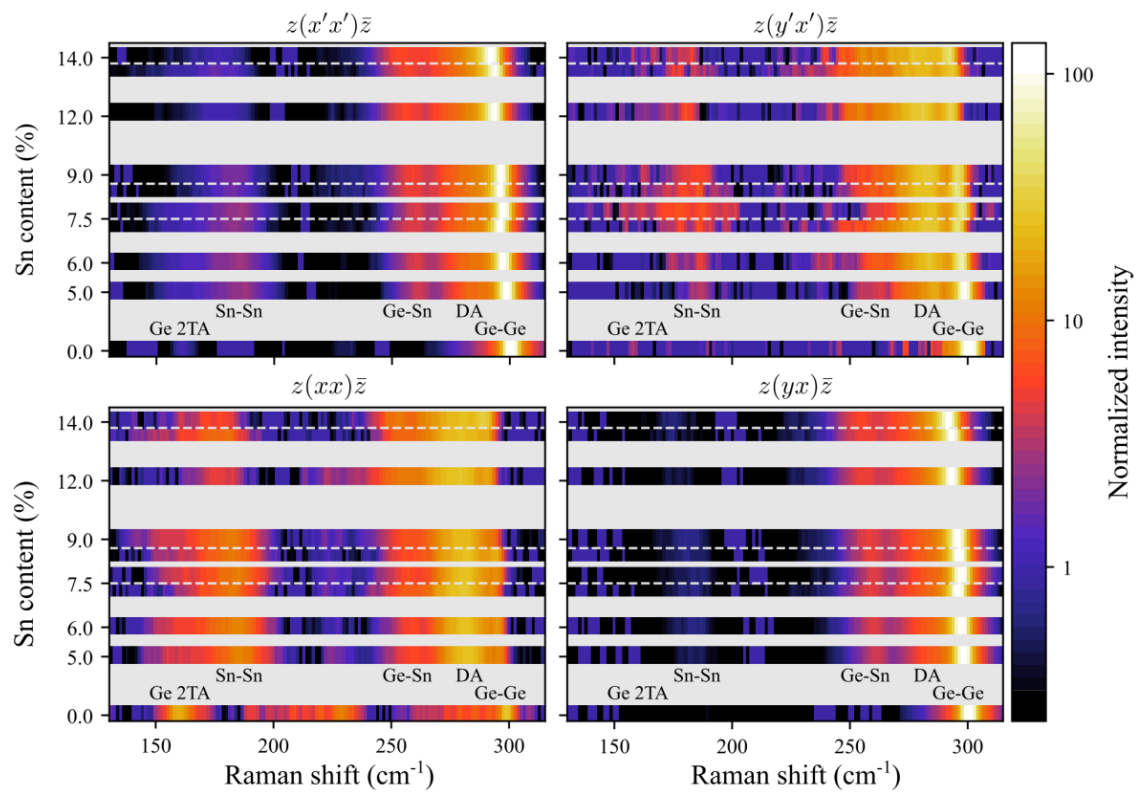

FIG. 6. Color maps of spectra for $Ge_{1-x}Sn_x$ with Sn content up to 14at.%, and Ge, for different polarization configurations. The spectra are normalized to the integrated intensity in the range 100-305 $cm^{-1}$. The intensity of the color scale is logarithmic. Gray areas are left for values of x for which no sample is available.

Having "tagged" all the spectral features observed on high and low content alloy, now the impact of Sn content on the intensity and energy of the polarization-resolved modes is investigated. To this purpose, in Fig. 6, the Raman spectra of Ge and $Ge_{1-x}Sn_x$ samples with Sn contents from 5 at.% to 14 at.% under different polarization configurations are presented as color maps. As a general trend, all the spectral features shift toward lower energies as the Sn content increases; for example, the energy of the Ge-Ge mode shifts from approximately 300 $cm^{-1}$ to approximately 291 $cm^{-1}$ by increasing the Sn content from 0 to 14 at.% Sn alloys. As discussed in the following, the residual heteroepitaxial strain is responsible for the deviation of the observed trend from a typical monotonic shift. In the previous section is remarked that the Ge-Ge, Ge-Sn, and Sn-Sn peak positions are almost independent of the polarization for all the Sn contents, while for DA, a clear difference in the peak position appears for the LO component and the ($A_{1g}$+$E_g$) component.

In Fc, the Ge-Ge and DA(LO), DA($E_g$) peak positions as a function of the Sn content are shown (for all the other peaks Appendix E). The peaks shift roughly parallel to each other with the Sn content. A linear regression $\omega_i(x,\varepsilon) = \omega_{i,0} + a_i x + b_i \varepsilon$, taking into account also the strain, yields the coefficient $a$ to be (-70±8) cm$^{-1}$, (-82±9) cm$^{-1}$ and (-78±13) cm$^{-1}$, respectively. The coefficient $b$ is 0 for all except the Ge-Ge mode. The full table for the coefficients is found in Appendix E. The data reported in Fc are chosen to represent the "pure" symmetry of LO and $E_g$. The cases with a combination of different symmetries are not reliable as the peak position is the "center of mass" of the convolved modes.

The presence of well-separated peaks for DA(LO) and DA($E_g$) at all the compositions supports the finding of the previous section, indicating that the distorted local environment is present even at lower Sn content, where a "shoulder" is not observed in the simulation of the ~6% alloy in FIG. 4. Interestingly, the coefficient $a$ is also an indicator of how much the incorporation of Sn modifies the interaction of a specific pair of atoms. The similar values for DA and Ge-Ge modes suggest that the local environment giving rise to the mode is indeed similar and thus similarly affected by the Sn. The Ge-Sn vibration mode is instead less sensitive to the Sn ($a$=-52±7 cm$^{-1}$). This can be explained considering that, as soon as the Sn is present, the pair Ge-Sn is less perturbed by the addition of more Sn atoms, and only its amplitude increases. As the Sn-Sn and 2TA modes are not strong enough at all polarizations and compositions, no conclusion can be drawn on their trend.

The insight gained through polarization-dependent measurements allows us to benchmark the results with literature data. These values are summarized in Appendix E, using data from Refs. [21,28,36–38]. A good agreement is found for both peak position and Sn-shift coefficient for Ge-Ge and Ge-Sn peaks. As expected, the use of polarized spectra is particularly relevant for the convolved DA mode, where an agreement is found with unpolarized data when compared with the $z(x'x')\bar{z}$ configuration that has all the representations active. For the Sn-Sn mode, the slight difference in peak position can be due to the fit of the two components.

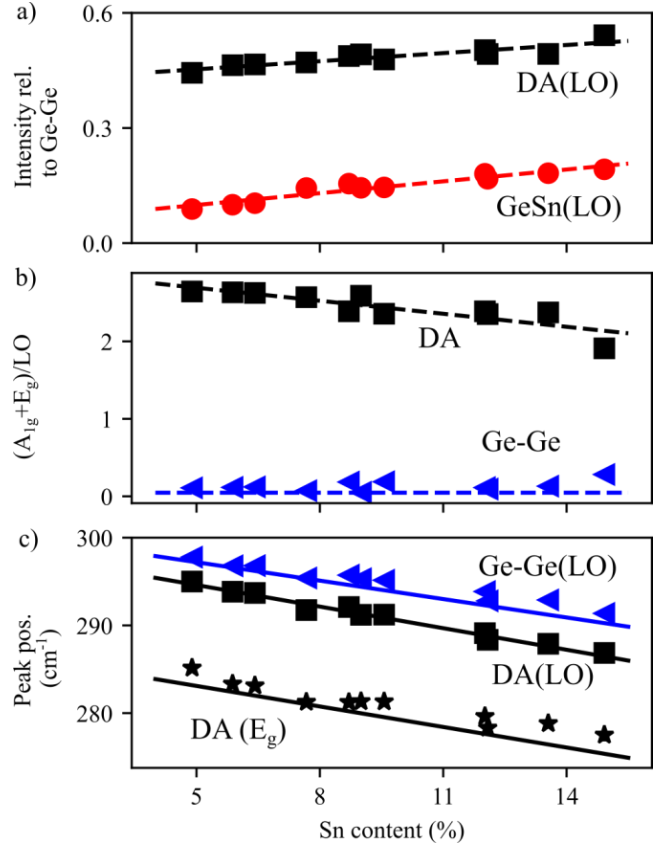

FIG. 7. Dependence of the peaks on the Sn content. a) Intensity (amplitude) of the DA and Ge-Sn peaks relative to the Ge-Ge peak, for the LO symmetry ($z(yx)\bar{z}$ configuration). b) Intensity of the $A_{1g}+E_g$ components relative to LO, for DA and Ge-Ge peaks, from $z(xx)\bar{z}$ and $z(yx)\bar{z}$ configurations, respectively. The intensity $I_k$ is calculated as $I_k/\Sigma I_k$, where k is GeGe, $Ge_{1-x}Sn_x$, DA. In panels a) and b), dashed lines are guides for the eyes. c) Peak positions for Ge-Ge ($z(yx)\bar{z}$ polarization) and DA, in the two components $E_g$ and LO (respectively from $z(y'x')\bar{z}$ and $z(yx)\bar{z}$ configurations). Solid lines are the result of the fit as described in the text.

The intensity of the Ge-Sn mode for the LO symmetry (Fa), as expected, increases with the number of Sn atoms. The DA(LO) component is also expected to increase, since more atoms will be in the distorted environment described in the previous section. Interestingly, the DA(LO) increases faster than the disorder-related DA($A_{1g}+E_g$), as shown in Fig. 7(b), indicating that the effect of the distorted environment on the spectrum is stronger than the effect of the disorder. By comparison, the main Ge-Ge peak has a very weak $A_{1g}+E_g$ component, which remains constant at all compositions. Indeed, the presence of a stronger DA(LO) mode at higher Sn content is consistent with the DFT calculation, where the simulated 6 at.% Sn alloy does not have an evident shoulder as for the case of approximately 19 at.% Sn. Again, this can confirm the onset of a short-range order, which is found to be dominant above 10 at.%.

## IV. Conclusion

In conclusion, polarized Raman spectroscopy and DFT numerical simulations provide insight into the role of Sn atoms on the vibrational properties of epitaxial $Ge_{1-x}Sn_x$ layers. These findings advance the understanding of the vibrational properties in $Ge_{1-x}Sn_x$ alloys, particularly with regard to the impact of local ordering of the different atomic species beyond simplified models.

In the whole investigated composition range from 5 at.% to 14 at.% Sn content, in the region of the main peak (approximately 300 cm$^{-1}$) the following features are identified: a shoulder with different symmetry from the main peak (approximately 280 cm$^{-1}$), and a shoulder or peak with the same symmetry of the main peak (approximately 290 cm$^{-1}$). The first one is assigned to disorder-activated scattering, while the second is assigned to Ge-Ge pairs whose bond is affected by local strain and lattice deformations caused by Sn atoms or clusters. The main peak is assigned to Ge-Ge pairs whose bond is least affected by local distortions.

The fact that different local environments can give rise to separate features has the potential to facilitate the development of methods for investigating the atomistic nature of the alloy, which could be utilized for monitoring the quality and characteristics of deposited layers using lab-scale Raman spectroscopy. At the same time, the role of these multiple environments can be very relevant in the electronic states, and thus in the optical properties of the alloy. Finally, the method presented here can be applied for the ternary silicon-germanium-tin alloy, to elucidate the effect of Sn on the local ordering.

## Acknowledgement

S.C. and T.L. are supported by the Air Force Office of Scientific Research under Award No. FA9550-19- 1-0341. This work was supported in part by high-performance computer time and resources from the DoD High Performance Computing Modernization Program.

## Appendix A: Samples description

The main properties of the samples are reported in Table II.

The Sn atomic content, $x$, the residual in-plane strain, $\varepsilon_\parallel$, and the degree of strain relaxation, $R$, were measured at room temperature by high-resolution X-Ray Diffraction (XRD) Reciprocal Space Mapping (RSM) using the 004 and 224 Bragg reflections. The relaxation is calculated as $R=(a-a_{sub})/(a_x-a_{sub})$, where: $a$ is the measured in-plain lattice constant of the strained layer, $a_x$ is cubic lattice constant for the alloy at $x$ Sn content, and $a_{sub}$ is lattice constant of the (virtual) substrate. The Sn content was extracted also by Rutherford backscattering spectroscopy (RBS).

*Table II Sample list and main properties.*

| Sample name | Thickness (nm) | Sn content XRD/RBS (%) | Strain (%) | Relaxation (%) |
|---|---|---|---|---|
| A5 | 770 | 4.9/5 | -0.13 | 77 |
| A6 | 660 | 6.4/6 | -0.16 | 81.3 |
| A8 | 380 | 5.9/7.7 | -0.13 | 84 |
| B7 | 770 | 7.8/7.3 | -0.32 | 80.5 |
| B8 | 420 | 8.7/8.5 | -0.29 | 75 |
| B9 | 421 | -/9 | -0.32 | |
| B11 | 442 | 9.6/11 | -0.33 | 76 |
| B12 | 700 | 12.1/12 | -0.32 | 80 |
| C12 | 290 | 12.0/12 | -0.62 | 61 |
| C13 | 350 | 13.5/13.6 | -0.55 | 70 |
| C14 | 470 | 14.9/14 | -0.51 | 76 |
| Ge | 2000 | 0 | +0.2 | 104 |

**Appendix B: Angular dependence of polarization**

In group IV semiconductors with diamond crystal structure, the point group is $O_h$. In Fig. 8 and Table III the angular dependence of the mode of possible Raman-active representations ($A_{1g}$, $E_g$, $T_{2g}$(LO)) of this point group are reported, considering the experimental geometry described in the inset of Fig. 8. The laser is incident along the growth direction *[001]* (*z*). The sample is rotated by an angle *β*, with *β=0* is the direction of the fixed analyzer. The laser polarization can be switched between a direction parallel and perpendicular to the analyzer. Considering the crystal directions *x=[100], y=[010], x'=[110], y'=[1$\bar{1}$0]*, the samples are mounted so that *x'* is aligned to *β=0*.

The $T_{2g}$(LO) mode has a distinctive pattern whose maximum intensity is equal for both configurations. In a parallel configuration, the intensity of the $T_{2g}$(LO) mode has its maximum for *β=0°, 90°, 180°* and *270°*, in perpendicular configuration for *45°, 135°, 225°* and *315°*. The $A_{1g}$ mode is forbidden in perpendicular polarization, while in parallel polarization its intensity is independent of $\beta$. $E_g$ mode intensity is more complex but clearly distinct from LO and $A_{1g}$.

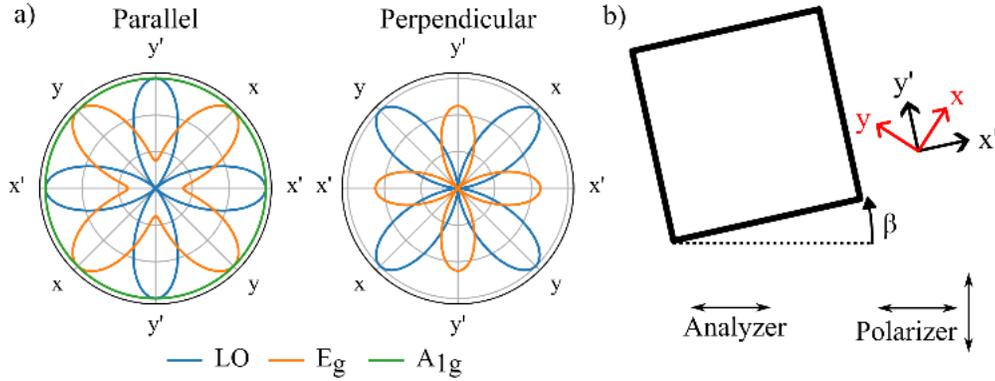

FIG. 8. Polar plots of the calculated intensity of Raman scattering for representations $A_{1g}$, $E_g$, $T_{2g}$(LO), as a function of the angle $\beta$. The inset illustrates the geometry.

Table III. Angular dependence for modes of different symmetry for parallel and cross configuration of the incident and scattered polarization. The azimuth $\beta$ of the sample is defined with respect to the [110] crystal axis.

| Representation | Parallel | Perpendicular |
| --- | --- | --- |
| $A_{1g}$ | 1 | 0 |
| $E_g$ | $\dfrac{(3\sin(2\beta)+1)^2}{4} + 3\cos^4\left(\beta+\dfrac{\pi}{4}\right)$ | $3\cos^2(2\beta)$ |
| $T_{2g}$(LO) | $\cos^2(2\beta)$ | $\sin^2(2\beta)$ |

## Appendix C: Peak fitting

For quantitative analysis of the peaks, we used exponentially modified Gaussian (EMG) lineshapes to reproduce the Ge-Ge mode asymmetry. At the same time, the Ge-Sn, DA peaks and the 2TA Ge mode and Sn-Sn modes were fitted using the Gaussian function. The fitting of the group around the main peak (Ge-Ge, DA and Ge-Sn peaks) is shown on Fig. 9 and the region below 220 cm$^{-1}$ (2TA and Sn-Sn) on Fig. 10. 2TA peak in Ge$_{1-x}$Sn$_x$ samples was fitted based on the reference position and width obtained on Ge spectra (Figs. 9(a) and 9(b))

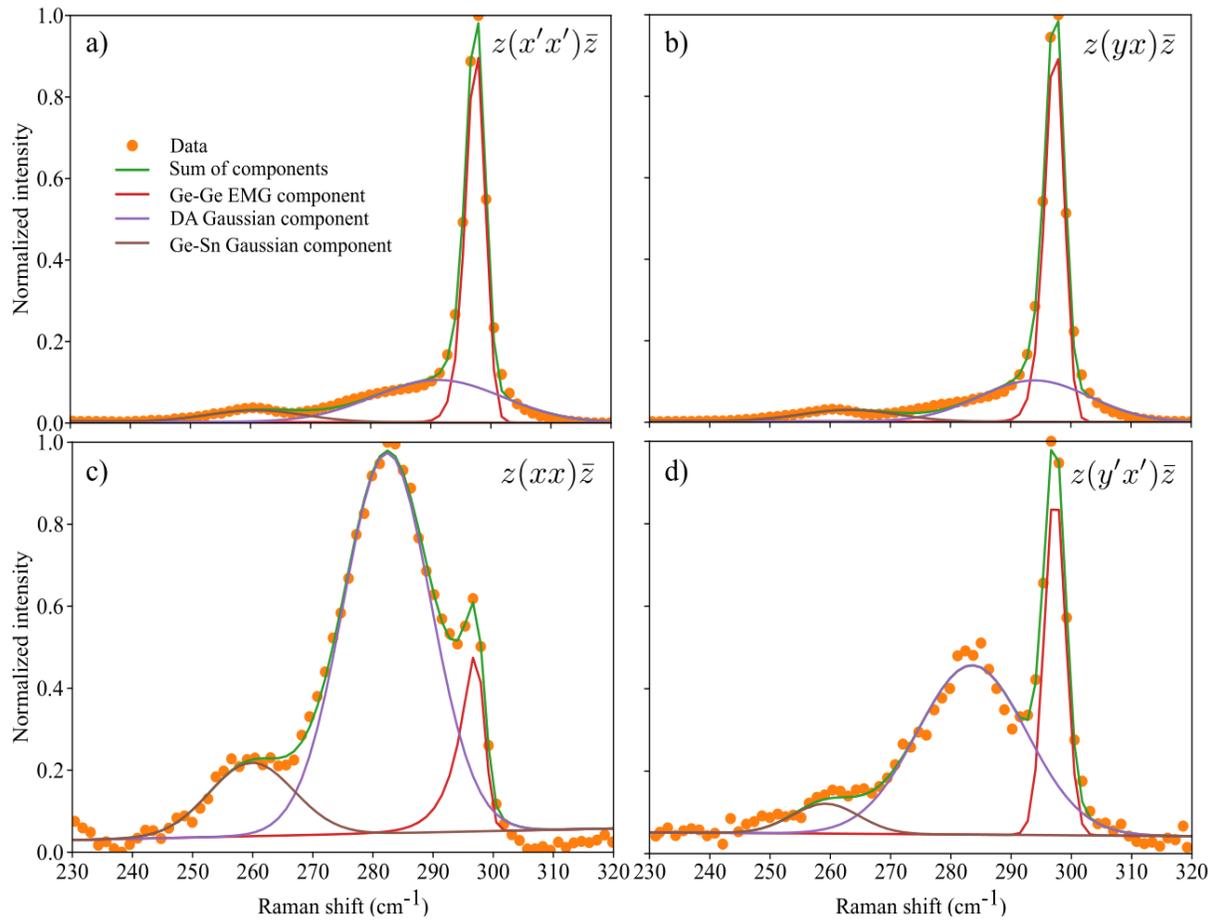

FIG. 9. Fitting of Ge-Ge, DA and Ge-Sn peaks of sample with 6% Sn in polarized configurations: *a)* z(x'x')z̄, *b)* z(yx')z̄, *c)* z(xx)z̄, *d)* z(y'x')z̄.

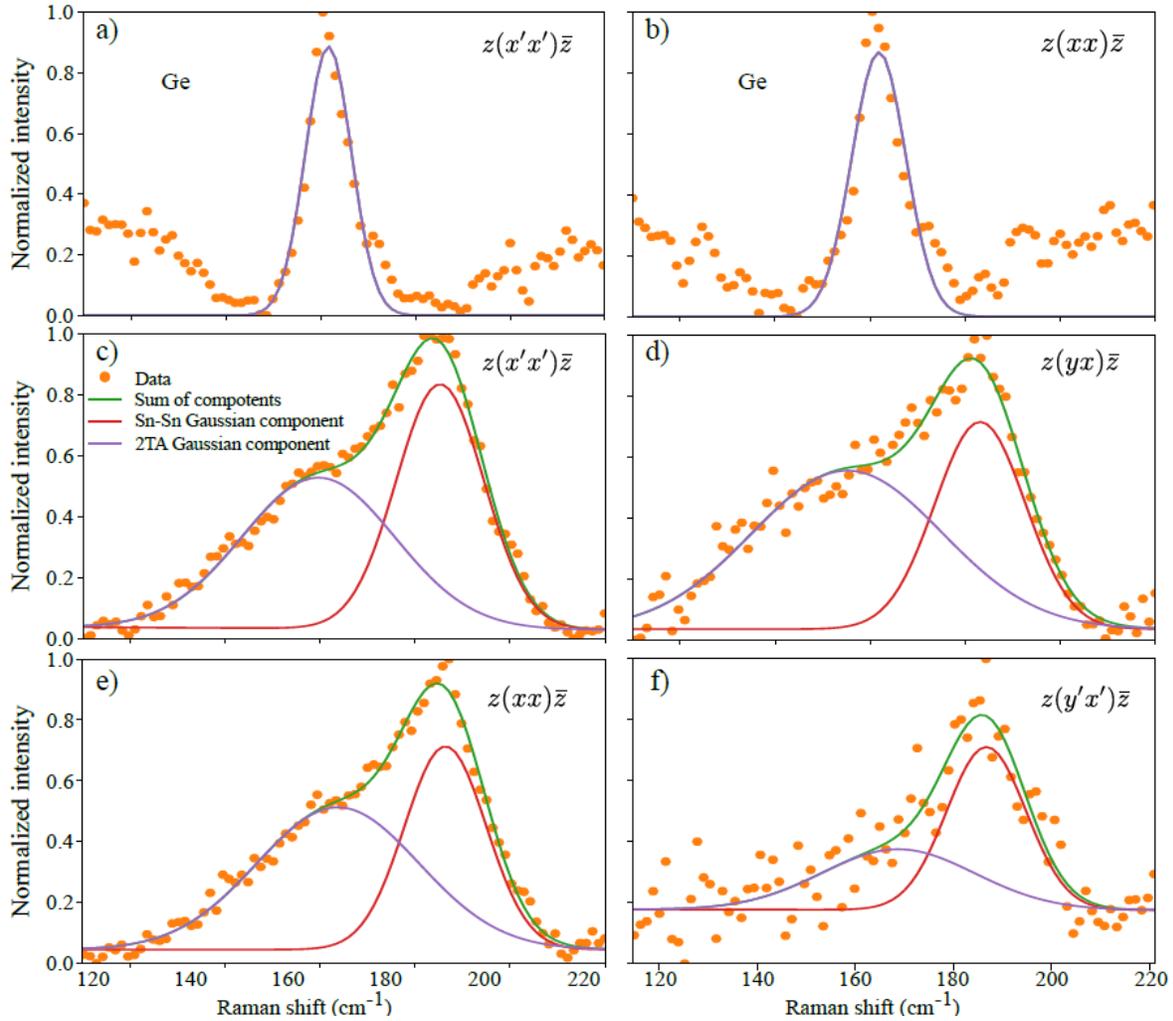

FIG. 10. Fitting of 2TA peak of Ge in polarized configurations: a) $z(x'x')\bar{z}$, b) $z(xx)\bar{z}$, 2TA and "Sn-Sn" peaks of sample with 6% Sn in polarized configurations: c) $z(x'x')\bar{z}$, d) $z(yx)\bar{z}$, e) $z(xx)\bar{z}$, f) $z(y'x')\bar{z}$.

## Appendix D: Simulation of low-energy peaks

Figure 11 shows the simulated displacement vectors for modes in the range of the "Sn-Sn" peak. Alloys with Sn content of 18.75 at.% and different ordering configurations are considered. It can be seen that these modes are not associated only to Sn-Sn pair vibrations, but also involve movement of Ge atoms, as discussed above (see also Fig. 4).

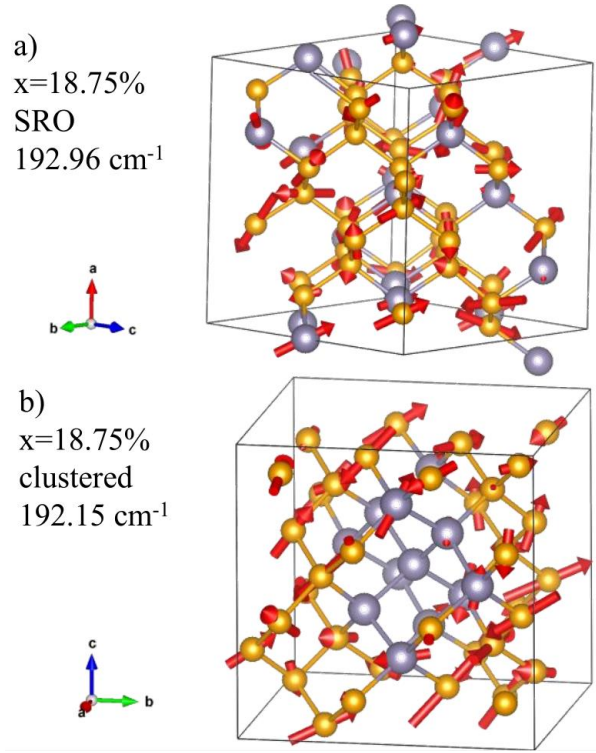

*FIG. 11. Displacement vectors for the commonly assumed "Sn-Sn" SRO (a) and clustered alloy (b) models, with x~19%. Sn atoms are represented as gray balls, Ge atoms are orange.*

## Appendix E: Dependence of peak positions on Sn content

Figure 12 shows the evolution of peaks' position for all the polarization configurations as a function of content.

Table IV gives the parameters extracted from linear regression as a function of Sn content $x$ and biaxial strain $\varepsilon$: $\omega_i(x,\varepsilon) = \omega_{i,0} + a_i x + b_i \varepsilon$. We have excluded the cases where the peaks are too weak for a reliable assessment.

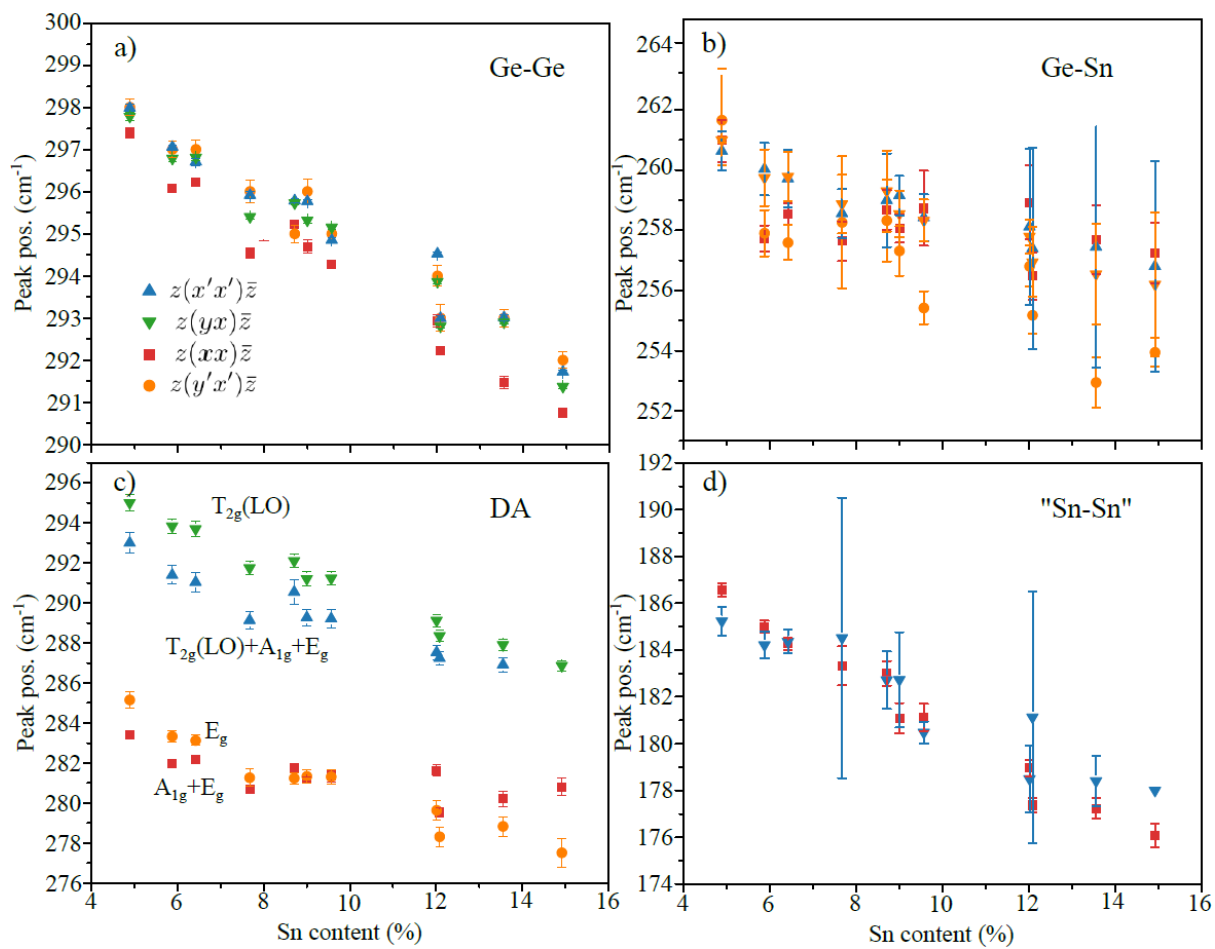

FIG. 12. Peak positions as a function of Sn content for a) Ge-Ge peak, b) Ge-Sn peak, c) DA peak, d) Sn-Sn peak.

*Table IV. Results of the linear regression of the peak positions as a function of Sn content. The representations active in every polarization configuration are reported. Literature values are also reported.*

| Polarization | Mode | $\omega_0$ – energy at x=0, $\varepsilon$=0 (cm$^{-1}$) | (literature) | $a$ – Sn shift (cm$^{-1}$) | (literature) | $b$ – strain shift (cm$^{-1}$) |
|---|---|---|---|---|---|---|
| $z(x'x')\bar{z}$ | Ge-Ge | 301.0±0.3 | 300.8 | -77±5 | -70 [21,36,38] | -420±110 |
| $A_{1g}+E_g+T_{2g}(LO)$ | DA | 294±1 | 291.3±0.7 [38] | -53±20 | -49 ± 7 [38] | -40±400 |
|  | Sn-Sn | 189±1 | 186-188 [37,38] | -83±10 |  | -140±90 |
| $z(xx)\bar{z}$ | DA | 283.9±0.8 | 287.9 [38], 285.8 [28] | -37±16 |  | -300±310 |
| $A_{1g}+E_g$ | Sn-Sn | 191.6±0.6 | 186-188 [37,38] | -120±12 |  | -320±240 |
| $z(yx)\bar{z}$ | Ge-Ge | 300.7±0.4 |  | -70±8 |  | -250±150 |
| $T_{2g}(LO)$ | DA | 298.7±0.4 |  | -82±9 |  | -50±170 |
|  | Ge-Sn | 262.8±0.4 | 262 [38], 263 [37] | -52±7 |  | -160±140 |
| $z(y'x')\bar{z}$ | Ge-Ge | 300.8±0.4 |  | -69±7 |  | -250±140 |
| $E_g$ | DA | 287±0.7 |  | -78±13 |  | -210±260 |